# Once highly productive, forever highly productive? Full professors' research productivity from a longitudinal perspective

Marek Kwiek[1,2,3] · Wojciech Roszka[2,4]

**Abstract**
This longitudinal study explores persistence in research productivity at the individual level over academic lifetime: can highly productive scientists maintain relatively high levels of productivity. We examined academic careers of 2326 Polish full professors, including their lifetime biographical and publication histories. We studied their promotions and publications between promotions (79,027 articles) over a 40-year period across 14 science, technology, engineering, mathematics, and medicine (STEMM) disciplines. We used prestige-normalized productivity in which more weight is given to articles in high-impact than in low-impact journals, recognizing the highly stratified nature of academic science. Our results show that half of the top productive assistant professors continued as top productive associate professors, and half of the top productive associate professors continued as top productive full professors (52.6% and 50.8%). Top-to-bottom and bottom-to-top transitions in productivity classes occurred only marginally. In logistic regression models, two powerful predictors of belonging to the top productivity class for full professors were being highly productive as assistant professors and as associate professors (increasing the odds, on average, by 179% and 361%). Neither gender nor age (biological or academic) emerged as statistically significant. Our findings have important implications for hiring policies: hiring high- and low-productivity scientists may have long-standing consequences for institutions and national science systems as academic scientists usually remain in the system for decades. The Observatory of Polish Science (100,000 scientists, 380,000 publications) and Scopus metadata on 935,167 Polish articles were used, showing the power of combining biographical registry data with structured Big Data in academic profession studies.

✉ Marek Kwiek
kwiekm@amu.edu.pl

Extended author information available on the last page of the article

## Introduction

This study explores persistence in research productivity at the individual level over academic lifetime. We examine the trajectories of the academic careers of 2326 Polish full professors, including their lifetime biographical histories and their lifetime publication histories. We studied the dates of their academic promotions and their publication output (79,027 articles) between promotions over a 40-year period across 14 science, technology, engineering, mathematics, and medicine (STEMM) disciplines. Our focus was on transitions between productivity classes throughout the professors' academic careers, from the assistant professor stage to the full professor stage.

We hypothesized that the current placement of full professors in the productivity classes of top, middle, and bottom (i.e., top 20%, middle 60%, and bottom 20% of scientists in prestige-normalized productivity in each discipline) corresponds, to some degree, to their placement in productivity classes at earlier stages of their careers. We speculated that current highly productive full professors could have also been highly productive associate professors and highly productive assistant professors earlier in their careers.

Our starting point was the current distribution of full professors by productivity classes in the 4-year period from 2014 to 2017. They were classified as either highly productive, average productive, or low productive. We then examined the productivity classes to which they could be retrospectively assigned at earlier stages of their careers.

The guiding thread of the paper follows the central findings of studies of highly productive scientists and their attributes (e.g., Fox & Nikivincze, 2021; Yin & Zhi, 2017; Agrawal et al., 2017; Cortés et al., 2016; Abramo et al., 2017; Kwiek 2016). Our research addresses three parallel questions: To what extent does scientists' research productivity change over their academic lifetime? Have currently highly productive scientists always been highly productive? Is it rare for radical changes (moving up or down) in productivity class to occur over academic careers? Most productivity studies focus on the individual traits of highly productive scientists, and many combine individual and organizational (environmental) features (Fox & Mohapatra, 2007; Fox & Nikivincze, 2021). Our approach to productivity analysis is longitudinal, relative (class based), and prestige normalized:

- *longitudinal*: we trace the productivity of full professors in our sample for several decades (since they entered the higher education system)
- *relative*: we do not examine publication numbers but focus on productivity classes, retrospectively assigning individuals to classes and comparing scientists to their peers in their disciplines and career stages
- *prestige normalized*: more weight is given to articles in high-impact journals than to those in low-impact journals, recognizing the highly stratified nature of academic science, especially in STEMM disciplines

Beyond its scholarly implications (in theories of productivity), persistent high productivity over the course of individual scientists' careers has implications for hiring and promotion policies. How can high productivity be maintained in departments, institutions, and national systems when, as this research shows, there is only marginal mobility between the lowest and highest productivity classes?

In this study, the unit of analysis was the individual researcher, not the individual publication. Although we used a combination of administrative, biographical, and bibliometric data, our research was not bibliometric in nature and belongs to academic profession studies. It was not possible to perform lifetime retrospective analyses of individual scientists without having full access to raw bibliometric metadata for all publications by all individual Polish scientists in the past 50 years. It was not possible to construct retrospective productivity classes for all scientists by discipline, career stage, and selected periods between promotions without having access to each scientist's global publication metadata - without the ability to collect structured Big Data from Scopus, a commercial bibliometric database. Our study provides an example of combining structured Big Data and national registry data to conduct detailed analyses of academic careers within a national (i.e., Polish) academic science system.

## Literature review

### High research productivity

For at least half a century, the sociology of science and of academic careers has addressed the theme of inequality in academic knowledge production (Hermanowicz, 2012), as a small percentage of scientists "contribute disproportionately to the advancement of science and receive a disproportionately large share of rewards and resources needed for research" (Zuckerman, 1988: 526). Within the Mertonian sociological tradition, the priority of discovery is important (Merton, 1973: 293), as one of the more salient motivations of scientists is "the desire for peer recognition" (Cole & Cole, 1973: 10). The scientific community, therefore, is not a "company of equals," and recognition of their work is "the only unambiguous demonstration that what [scientists] have done matters to science" (Zuckerman, 1988: 526). The recognition afforded by publications and citations translates into funds for further research, and the distribution of output, citations, academic awards, and research funding is highly stratified. At an institutional level, inequality and stratification are pervasive features of higher education systems, with "endless competition" between "status-seeking" institutions (Taylor & Cantwell, 2019); at the individual level, research powerfully segregates the academic profession, and rewards are "distributed in a highly stratified fashion" (Marginson, 2014: 107).

In any system of academic science, a small number of scientists publish the majority of papers and attract the most citations (Abramo et al., 2009; Ruiz-Castillo & Costas, 2014; Stephan, 2012). In every scientific community, highly productive scientists hold prestigious academic positions and are responsible for shaping the identity of academic disciplines (Cortés et al., 2016). How do highly productive scientists emerge in higher education? Scientific productivity is believed to result from (1) individual attributes, (2) organizational attributes (academic environment), and (3) features of the national academic science system, within which the allocation of rewards and recognition of research achievements play important roles. Science is a complicated social institution, and scientists must be systemically supported within their national edifice of science to sustain high productivity over time. Indeed, the efficient operation of science depends on how it "divides up the rewards and prizes it offers for outstanding performance, and structures opportunities for those who hold extraordinary talent" (Cole & Cole, 1973: 15).

Access to resources is enjoyed by those held in high esteem within the scientific community, who are strongly motivated to publish because scientific esteem "flows to those who are highly productive" (Allison & Stewart, 1974: 604). Highly productive scientists are those whose productivity persists over time (Abramo et al., 2017), they are the small group that maintains high productivity in their own work, supported or not by structural features of the science system, including mechanisms for accumulating advantages over time. Cumulative advantage is a broader process by which "small initial differences compound to yield large differences" (Aguinis & O'Boyle, 2014: 5). In science, consequently, the Matthew effect leads to inequalities in access to financial and nonfinancial rewards (Xie, 2014).

Historically, the sociology of science shows that scientific recognition is rooted almost exclusively in research production (Cole and Cole, 1973), and the reward system is structured to benefit scientists who best perform their role. In Merton's (1973: 297) words, "The institution of science has developed an elaborate system for allocating rewards to those who variously live up to its norms." In Merton's reputation- and resource-based model of scientific careers, new resources are not simply rewards for past high productivity but serve the primary function of stimulating future high productivity: "The scientific community favors those who have achieved significant success in the past" (DiPrete & Eirich, 2006: 282). In the past decade, studies have looked intensively at high research productivity (e.g., Yair et al., 2017; Aguinis & O'Boyle, 2014; Agrawal et al., 2017; Abramo et al., 2017; Yin & Zhi, 2017; Piro et al., 2016; Kwiek, 2016, 2018). Most recently, Fox and Nikivincze (2021) studied highly prolific scientists from a social-organizational perspective that examined both individual characteristics and departmental features. They identified three predictors of high productivity: rank, collaborative span, and a favorable work climate (*work climate* being a perceived departmental atmosphere that stimulates [or stifles] performance) (Fox & Mohapatra, 2007). Abramo et al. (2017), whose research most nearly resembles ours, examined the research performance of all Italian professors in the sciences over three consecutive 4-year periods (2001–2012). Their analyses demonstrate that 35% of top scientists retain their high productivity for three consecutive periods, and 55% do so for two periods. Higher percentages of male scientists than female scientists keep their stardom, with some cross-disciplinary differences (Abramo et al., 2017: 793–794). Our research differs from theirs in time span (lifetime vs. 12 years), sample selection (full professors vs. all professors), and methodology (three productivity classes vs. top performers and unproductive scientists; prestige-normalized productivity vs. Fractional Scientific Strength).

Several key theories have emerged from the sociology and economics of science to explain dramatic differences in individual research productivity, which may be useful in studying the stratification of Polish scientists. The "sacred spark" theory (Cole & Cole, 1973) states that "there are important, predetermined differences among scientists regarding their ability and motivation for creative scientific research" (Allison & Stewart, 1974: 596). Highly productive researchers "are motivated by an inner drive to create science and by a pure love of the work" (Cole & Cole, 1973: 62). Productive scientists are a highly motivated group of researchers and have the necessary "capacity to work hard and persist in the pursuit of long-range goals" (Fox, 1983: 287). Stephan and Levin (1992: 13) hold a similar view, stating that "there is a general consensus that certain people are particularly good at doing science and that some are not just good but superb." Cumulative advantage theory (Merton, 1973) holds that productive scientists will be even more productive in the future, while low-productivity scientists become even less productive over time. "Scientists who are rewarded are productive, while scientists who are not rewarded become less productive" (Cole & Cole, 1973: 114). Finally, the utility maximization theory, which emerged from the economics of science, asserts that researchers reduce their research-oriented efforts with time because they believe that other tasks may be more personally rewarding for them. Discussing aging and productivity, Stephan and Levin (1992: 35) argue that "later in their careers, scientists are less financially motivated to conduct research" (see Kyvik, 1990). These three main theories of research productivity complement one another and apply to varying degrees to the academic profession in Poland (Kwiek, 2019: 27-32). The sacred spark and cumulative advantage theories explain high research productivity, while low productivity in Poland may be understood through both the cumulative advantage and utility maximization theories.

Admission to the class of the most productive scientists requires a strong research orientation and long hours spent on research (Kwiek, 2016, 2018) in addition to the innate capacities highlighted by the sacred spark theory and the prior achievements stressed in the cumulative advantage theory. A high proportion of the most productive scientists will always be among the most productive—regardless of circumstances, location in the system, age, and career stage—while only a small proportion of low-productivity scientists ever become highly productive as shown in this study. In the process of accumulating advantages, exceptional research productivity early in a career translates into new resources and rewards that make it easier to sustain high research productivity in subsequent years and decades. Research resources are not rewards for past productivity but are designed to stimulate the productivity of the most productive in the future: "The scientific community favors those who have achieved the most in the past in terms of the additional resources and attention they have enjoyed" (DiPrete & Eirich, 2006: 281–282).

## Full professorship: international insights

Research on full professors in academia is important to our study, as our sample is specific (including only full professors) and the topic is rarely studied due to the limited availability of data on academic ranks. Thus, we briefly discuss selected recent papers on full professors. In the US context, Yuret (2018) analyzed the "paths to success" in obtaining full professorships and found that promotion to full professorship was related to high mobility and short duration of PhD studies. Kolesnikov et al. (2018) studied full professors in two fields at 10 research-intensive universities to test the hypothesis that productivity is inversely correlated with impact. However, higher productivity led to lower impact in one field and to higher impact in the other.

The extant literature also includes studies in Israel and Norway. Weinberger and Zhitomirsky-Geffet (2021) examined diversity in scholarly performance among tenured professors at Israeli universities by distinguishing between high-, average-, and low-impact scholars. The results of their linear regression analysis show that women outperformed men in terms of scientific impact, and these differences in performance reveal that scholarly success and promotion to full professorship may not be fully determined by productivity (Weinberger & Zhitomirsky-Geffet, 2021: 2949). In Norwegian universities, Piro et al. (2016) studied the influence of prolific full professors on the citation impact of their university department. While productivity was skewed at the level of individuals, the influence of prolific professors on their departments' citation impact was modest.

Fox (2020: 1002) argues that gender predicts academic rank: "Women are less likely than men to hold higher ranks, and the gender disparity is especially apparent for the rank of professors." Regarding access to full professorships by gender, the recent evidence is inconclusive. Using multilevel logistic regression, Marini and Meschitti (2018) found that men had a 24% higher probability of being promoted in Italy than women with the same scientific output. Madison and Fahlman (2020) analyzed all promotions to full professorships in Swedish institutions, however, and found no bias against women attaining full professorships in relation to publication metrics.

Lerchenmueller and Sorenson (2018) studied gender gaps in career transitions in the life sciences in the USA and found that the gender gap largely emerges during the brief period when men and women move from working in another researcher's lab to leading their own (Lerchenmueller & Sorenson, 2018: 1015). In Germany, Lutter and Schröder (2016) show that women sociologists were likelier than men with the same number of publications to obtain a full professorship. Among men, the strongest predictor of success was publication in Social Science Citation Index (SSCI) journals. Among women, by contrast, the strongest predictor was the accumulated number of academic awards. Drawing upon the profiles of 2,528 scholars, Habicht et al. (2022) examined gender differences in obtaining tenured psychology professorships in Germany, and they reject the female devaluation theory, which holds that women's career achievements are devalued in relation to those of men.

Puuska (2010) focused on ranks, productivity, and various types of publication by 1,417 Finnish professors. The results show that full professors were the most productive; the principle of "the higher, the more productive" applied to all academic ranks (Puuska, 2010: 428–430). In several male-dominated fields, female full professors were more productive than male full professors, which may indicate that, in those fields, only exceptionally productive women win full professorships (Puuska, 2010: 435). Abramo et al. (2011) examined the links between individual productivity and academic rank among Italian university researchers. The results reveal a uniform productivity distribution across the ranks that only slightly favors full professors. Full professors exhibited the highest productivity, but "top scientists" (i.e., the upper 10%) were evenly distributed among the three ranks (Abramo et al., 2011: 927). Aksnes et al. (2011) show that Norwegian full professors have a lower-than-average citation index despite their high productivity index; by far, the highest indexes in both categories were obtained by postdocs (Aksnes et al., 2011: 632). Those authors also studied the impact of mobility on productivity by comparing mobile and non-mobile Norwegian scientists, finding that the mobile full professors were the most productive group (Aksnes et al., 2011: 219). Finally, in the US context, Fox and Nikivincze (2021: 1250) show that, compared with the rank of assistant professor, the rank of full professor was a strong and positive predictor of being highly prolific.

**Research questions and hypotheses**

The six research questions in Table 1 were based on selected findings in the previous studies mentioned in the "High research productivity" section (RQ1, RQ2, RQ4, and RQ6) and in the "Full professorship: international insights" (RQ3 and RQ5). The hypotheses pertain to the persistence of high productivity (H1) and low productivity (H2) over time; the persistence of high productivity at the beginning and towards the end of academic careers (H4); disciplinary differentiation (H3) and gender differentiation (H5) in mobility between productivity classes; and (H6) the role of past productivity class membership in estimating (via logistic regression analysis) the odds ratio of belonging to top productivity classes. An overarching research question concerns changes in productivity from a life-cycle perspective: have currently top-performing full professors always been top performing, and have currently low-performing full professors always underperformed?

Table 1 Research questions, hypotheses, and summary of support

| Research questions | Hypotheses | Support |
|---|---|---|
| RQ1. What is the relationship between currently high productivity and high productivity in the two earlier stages of an academic career? | **Persistence of *high* productivity over time**<br>H1: Currently highly productive full professors were, in a significant proportion, highly productive associate professors, and highly productive associate professors were, in a significant proportion, highly productive assistant professors | Supported |
| RQ2. What is the relationship between currently low productivity and low productivity in the two earlier stages of an academic career? | **Persistence of *low* productivity over time**<br>H2: Currently low-productive full professors were, in a significant proportion, low-productive associate professors, and low-productive associate professors were, in a significant proportion, low-productive assistant professors | Supported |
| RQ3. What is the relationship between productivity trajectories during a life cycle and academic disciplines? | **Disciplinary differentiation**<br>H3: Mobility between productivity classes varies by discipline | Supported |
| RQ4. What is the relationship between current productivity and productivity at the beginning of an academic career? | **Persistence of productivity throughout the academic careers**<br>H4: Current full professors belong, in a significant proportion, to the same productivity class at the beginning and at the end of their academic careers | Supported |
| RQ5. What is the relationship between productivity trajectories over a life cycle and gender? | **Gender differentiation**<br>H5: Mobility between productivity classes varies by gender | Supported |
| RQ6. What is the effect of past productivity on membership in top productivity classes in the context of the joint effects of other variables? | **Model approach to top productivity classes: logistic regression analysis**<br>H6: Past membership in top productivity classes significantly increases the odds ratio estimates of memebership in top productivity classes | Supported |

# Context, data, methods, and sample

## The national context

With 1,218,000 students, Poland's higher education system employs 88,416 full-time academics (48.04% women) distributed among the major ranks as follows: 8,990 full professors (27.49% women), 17,303 associate professors (39.84% women), and 38,978 assistant professors (50.36% women). Regarding promotion to higher ranks in the higher education sector, there were 597 new full professors in 2021 (220 women; 36.85%) and 490 new associate professors (212 women; 43.27%); furthermore, 3,431 doctoral degrees were awarded (1752 women; 51.06%) (GUS, 2022: 30–38). Publications come predominantly from universities in several major academic cities, with the bulk of internationally visible academic knowledge production coming from the 10 research-intensive Excellence Initiative—Research Universities (IDUB) institutions selected to receive additional funding in 2020–2026. The growth in globally indexed publications has been substantial in the past decade, increasing about 100% (from 31,707 in 2010 to 62,131 in 2021).

More than 90% of new hires are in-house hires (i.e., academics with doctoral degrees from the same institution); cross-institutional mobility is of marginal importance, and promotion to higher ranks is almost exclusively within the same institution. Salaries and workloads are similar across the system and regulated at a national level. All academics in the three ranks are expected to be equally involved in teaching and research. Promotion to the ranks of associate and full professor are nationally governed, related to degrees obtained (the habilitation and professorship degrees), and based on research achievements assessed by peer committees. Full professorship, the crowning achievement in an academic career, is desired by many but available to few (about 600 new professorships annually in the past few years). There are no institutional or national limits on the number of new full professorships, which are awarded on the basis of successfully passing rigorous, research-based national promotion procedures. Only one dimension of the academic career matters in promotions: research output since earning the habilitation degree. All assistant professorships are tenure-track positions, and tenure is granted to associate professors (upon obtaining the habilitation degree), with long-term job contracts and job stability for the vast majority of academics. In the past few years, expectations of publication in globally indexed journals have notably risen (although they have always been high in STEMM disciplines). Reward structures are similar across the system, with higher ranks offering higher salaries and greater participation in university self-governance.

## Dataset and sample

The data used in this study were collected from a national administrative and biographical register of all Polish scientists ($N=99,935$) and from the Scopus bibliometric database (2009–2018, $N=380,000$ publications). The final number of articles was 158,743, and they were published by 25,463 unique authors with Polish affiliations. The database was then enriched with publication metadata collected from Scopus, which were obtained through a collaboration agreement with the ICSR Lab, which is a cloud-computing platform provided for research purposes by Elsevier ($N=935,167$ articles from 1973–2021 by authors with Polish affiliations). We used information about the entire academic output of individual authors based on their Scopus Author IDs in the database. Our final sample included full professors in 14 STEMM disciplines ($N=2326$) who authored or co-authored 79,027 articles.

## Defining academic disciplines and academic age

We defined individual attributes for the sample of 23,543 scientists in all academic positions and disciplines, including every full professor in the 14 STEMM disciplines in our final sample. In the All Science Journal Classification (ASJC) system of disciplines used in Scopus, a journal publication has one or multiple disciplinary classifications. The dominant discipline of each full professor was determined based on all publications (type: article) included in their individual publication portfolios for the period from 2009–2018 (the modal value is the most frequently occurring value). When there was no single value, the dominant discipline was randomly selected from among the most frequently occurring disciplines.

Our dataset included the professors' year of birth and the year in which every full professor achieved three scientific degrees—the doctoral degree, habilitation degree, and professorship—which were used as proxies for assistant, associate, and full professor, respectively. The three degrees are clear markers in their careers, and their details (date, title, institution, discipline, field, reviewers) are available in our dataset. Full professors entered the Polish equivalent of tenure-track positions when they obtained their doctoral degrees and received permanent employment (the equivalent of tenure) when they obtained their habilitation degrees. As an analytical approach, we chose a three-degree system rather than a multiple-rank system (with "university professors" and "ordinary professors") because the latter system is not consistently applied across all institutions. For an international audience, the best way to discuss promotions in rank in Poland is through the three-degree system, with clearly defined dates for each scientist and with nationally determined, research-based requirements.

We obtained the year of the first publication indexed in Scopus using the application programming interface (API) protocol, which is a set of programming codes that enable data transmission between one software product and another provided by Scopus. The gender of all scientists with at least a PhD degree is included in the data provided by the national registry of scientists, and in this study, it was treated as a binary variable.

## Full professors: discipline, age, and gender distribution

The distribution of our final sample was as follows: about three-fourths of full professors were men (see Table 2); about one-third worked in 10 research-intensive IDUB institutions. The three disciplines with the largest number of full professors were medical sciences (MED), agricultural and biological sciences (AGRI), and engineering (ENG). About half of all Polish full professors in our sample were publishing in these three disciplines. The largest share of female full professors in larger disciplines was in biochemistry (BIO), MED, and AGRI (about one-third). The lowest share was in physics and astronomy (PHYS) (5.5%), mathematics (MATH) (6.3%), and ENG (5.8%). About two-thirds were aged more than 60 years and about a half were aged from 65–70 years. In our sample, 16% were young (under 55 years) full professors, including 2% aged 40–44 years. The distribution by biological age and gender are presented in Fig. 1. The distribution of female scientists was equal across ages, while the distribution of male scientists was steeper. The distribution by age was more similar than expected. The gender distribution of the full professors in our sample was close to their gender distribution in the population over the past five years (GUS, 2022).

**Table 2** Structure of the sample of all Polish internationally visible university full professors by gender, age group, and STEMM discipline

|  |  | Female scientists | | | Male scientists | | | Total | | |
|---|---|---|---|---|---|---|---|---|---|---|
|  |  | n | row % | col % | n | row % | col % | n | row % | col % |
| Age groups | Total | 551 | 23.7 | 100.0 | 1775 | 76.3 | 100.0 | 2326 | 100.0 | 100.0 |
|  | up to 50 | 48 | 24.9 | 8.7 | 145 | 75.1 | 8.2 | 193 | 100.0 | 8.3 |
|  | 51—60 | 164 | 27.2 | 29.8 | 438 | 72.8 | 24.7 | 602 | 100.0 | 25.9 |
|  | 61—65 | 145 | 22.3 | 26.3 | 505 | 77.7 | 28.5 | 650 | 100.0 | 27.9 |
|  | 65–70 | 194 | 22.0 | 35.2 | 687 | 78.0 | 38.7 | 881 | 100.0 | 37.9 |
| IDUB | IDUB | 130 | 16.7 | 23.6 | 650 | 83.3 | 36.6 | 780 | 100.0 | 33.5 |
|  | Rest | 421 | 27.2 | 76.4 | 1125 | 72.8 | 63.4 | 1546 | 100.0 | 66.5 |
| Academic discipline | AGRI | 119 | 33.9 | 21.6 | 232 | 66.1 | 13.1 | 351 | 100.0 | 15.1 |
|  | BIO | 66 | 37.9 | 12.0 | 108 | 62.1 | 6.1 | 174 | 100.0 | 7.5 |
|  | CHEM | 41 | 25.2 | 7.4 | 122 | 74.8 | 6.9 | 163 | 100.0 | 7.0 |
|  | CHEMENG | 9 | 21.4 | 1.6 | 33 | 78.6 | 1.9 | 42 | 100.0 | 1.8 |
|  | COMP | 14 | 14.4 | 2.5 | 83 | 85.6 | 4.7 | 97 | 100.0 | 4.2 |
|  | EARTH | 13 | 11.3 | 2.4 | 102 | 88.7 | 5.7 | 115 | 100.0 | 4.9 |
|  | ENER | 6 | 19.4 | 1.1 | 25 | 80.6 | 1.4 | 31 | 100.0 | 1.3 |
|  | ENG | 18 | 5.8 | 3.3 | 292 | 94.2 | 16.5 | 310 | 100.0 | 13.3 |
|  | ENVIR | 57 | 35.6 | 10.3 | 103 | 64.4 | 5.8 | 160 | 100.0 | 6.9 |
|  | MATER | 37 | 23.1 | 6.7 | 123 | 76.9 | 6.9 | 160 | 100.0 | 6.9 |
|  | MATH | 9 | 6.3 | 1.6 | 133 | 93.7 | 7.5 | 142 | 100.0 | 6.1 |
|  | MED | 138 | 36.4 | 25.0 | 241 | 63.6 | 13.6 | 379 | 100.0 | 16.3 |
|  | PHARM | 14 | 66.7 | 2.5 | 7 | 33.3 | 0.4 | 21 | 100.0 | 0.9 |
|  | PHYS | 10 | 5.5 | 1.8 | 171 | 94.5 | 9.6 | 181 | 100.0 | 7.8 |

*Note*: STEMM disciplines included in the study: AGRI, agricultural and biological sciences; BIO, biochemistry, genetics, and molecular biology; CHEMENG, chemical engineering; CHEM, chemistry; COMP, computer science; EARTH, earth and planetary sciences; ENER, energy; ENG, engineering; ENVIR, environmental science; MATER, materials science; MATH, mathematics; MED, medical sciences; PHARM, pharmacology, toxicology, and pharmaceutics; and PHYS, physics and astronomy

## Methodological approach

### Constructing lifetime biographical and lifetime publication histories

The Laboratory of Polish Science database constructed and maintained by the authors includes the complete publication histories of all Polish scientists working in the higher education sector as of November 2017, holding at least a PhD degree, and having at least one publication in the Scopus database. The database includes the publication and citation metadata on all publications by each scientist in each stage of their scientific career. The database included data on 14,271 assistant professors, 7,418 associate professors, and 3,774 full professors in STEMM and non-STEMM disciplines.

However, we focused only on full professors, which enabled us to trace their individual biographical histories and individual publication histories in the earlier stages of their careers. Only full professors could be compared in three earlier stages. The analysis of full professors included a long period of scientific activities lasting several decades. We retrospectively examined the academic career classes of full professors who had been working for 20–40 years. The compilation of complete lifetime biographical histories (i.e., years of birth and years of subsequent academic promotions) and complete lifetime publication histories (i.e., detailed data on publications, collaborations, mobility, and citations), spanning entire academic careers, allowed us to retrospectively analyze the transitions between productivity classes over time of all full professors.

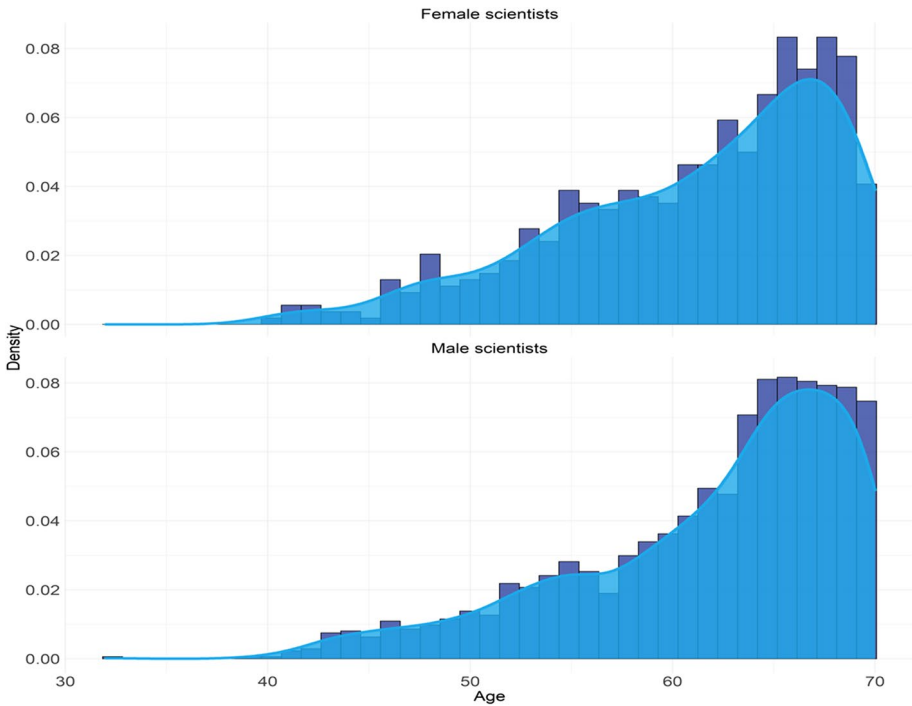

**Fig. 1** Distribution of biological age: kernel density plot, full professors in 14 STEMM academic disciplines combined, by gender

In this study, we applied a longitudinal approach to analyzing the transitions between the productivity classes of the full professors over their careers, from the year in which they received their PhD degrees to 2017. We analyzed the productivity of individual scientists as they aged and moved up the academic ladder. Each publishing scientist within their unique biographical history (based on dates) and unique publication history (based on publication metadata) was characterized by transitions between productivity classes compared with their peers in the same discipline and at the same career stage.

## Constructing prestige-normalized research productivity

The productivity of researchers at a given stage in their academic career was calculated as the number of all publications (publication type: article) produced in that stage divided by the number of years spent at the stage (and multiplied by 4 to maintain the comparability of productivity over 4-year reference periods). Productivity may vary during academic careers, with pre-promotion peaks and post-promotion pauses (Katz, 1973), so this approach reduced potential differences between the first years after each promotion (when productivity may decrease) and the years just before a new promotion (when productivity may increase). We divided the academic careers of the full professors into three stages based on distinct opening and closing dates (doctorate, habilitation, full professorship), and we constructed both lifetime productivity profiles and productivity profiles in their three distinct career stages. We used a full counting approach instead of a fractional counting approach in which single-authored and multiple-authored publications were counted equally. We used the prestige-normalized publication number rather than the raw publication number.

In prestige-normalized productivity, we combined the output indicator of research productivity with the indicator of scholarly impact on science (based on citations). Output indicators measure the knowledge produced, and impact indicators measure the ways in which scholarly work affects the research community (Sugimoto & Larivière, 2018: 1). The weight of an article depends on its position in the global hierarchy of academic journals. In our approach, articles published in journals with, on average, a high impact on the academic community captured through the proxy of average citation numbers were given more weight in calculating productivity than articles in low-impact journals because they required, on average, more scholarly effort to write and get published. Our approach to productivity recognizes the highly stratified nature of academic science, in which both the quantity of publications and their standardized quality are important.

Measuring journal prestige is closely related to the Polish system of evaluating scientists and scientific units and to the indicators used in the IDUB national excellence program. Articles in highly prestigious journals require, on average, a greater workload and have, on average, greater resonance in the world of science, as captured through citations. In Scopus, the prestige rank of a journal is determined annually by the journal's placement in the CiteScore ranking system, which is prepared annually for all journals indexed (e.g., 40,562 in 2022). Percentile ranks are based on values in a range from 1 to 99, in which the highest prestige is the 99th percentile. Highly prestigious journals in each field, with low acceptance rates, tend to be in the 90–99th percentile (*Higher Education* and *Studies in Higher Education* are in the 96th percentile of Scopus journals). Publications in more prestigious journals count more in productivity calculations compared with publications in less prestigious journals within each discipline.

In a non-normalized approach to productivity (full-counting), an article published in any journal would receive a value of 1. In contrast, in the prestige-normalized productivity approach, an article in a journal with a percentile rank of 90 received a value of 0.90, while an article in a journal with a percentile rank of 40 received a value of 0.40. Articles published in journals with percentile ranks of and below 10 received a value of 0.1. A prestige-normalized approach to individual research productivity allows for a fair measurement of scholarly effort in STEMM disciplines in which vertical journal stratification is a fact of life. Counting all publications in the same way would disregard individual scholarly efforts invested in research. Each discipline has specific highly competitive top-tier journals, and "the tyranny of the top five" (Heckman & Moktan, 2018) is applicable far beyond economics.

### Constructing academic career classes

This study draws upon the notion of climbing the academic ladder, which defines the decades-long academic careers of full professors. The current full professors (promotion date: full professorship awarded) were initially assistant professors (promotion date: doctoral degree awarded) and then associate professors (promotion date: habilitation degree awarded). They all remained for a specific number of years at previous stages of their academic careers. At each stage, they demonstrated specific productivity. We ranked all academics (segregated by discipline) by their 4-year prestige-normalized research productivity within specific career stages. For each full professor, we counted all articles published within the stages as defined by promotion dates: the first stage is between doctoral degree and habilitation degree (which we term *assistant professorship* to reflect internationally understandable career steps), the second stage is between the habilitation degree and the title of professor (which we term *associate professorship*), and the third stage is between the professorship title and 2017 (which we term *full professorship*). For instance, if a biographical history of full professor X shows that she obtained her doctoral degree in 1995, habilitation degree in 2002, and full professorship in 2012, then her assistant professorship stage was 1995–2001, associate professorship stage 2002–2011, and full professorship stage 2012–2017.

Central to our analysis was the current distribution of full professors by productivity classes in the 4-year period from 2014–2017. They were classified as either high productivity, average productivity, or low productivity full professors. We then examined the productivity classes to which they could be retrospectively assigned at earlier stages of their careers, that is, when they were assistant professors and associate professors.

We assigned seven academic career classes to each full professor (see Fig. 2): three productivity classes, two promotion age classes, and two promotion speed classes. The current and past productivity classes were the top, middle, or bottom—that is, the upper 20%, middle 60%, or lower 20%, respectively, in a prestige-normalized and discipline-normalized approach separately within each of the 14 STEMM disciplines. The promotion age classes were young, middle, or old associate professors and young, middle, or old full professors. That is, the upper 20%, middle 60%, or lower 20%, respectively, in terms of promotion age expressed in full years. The promotion speed classes were fast track, typical track, and slow track associate professor and fast track, typical track, and slow track full professor, that is, the upper 20%, middle 60%, and lower 20%, respectively, in terms of the transition time between subsequent promotions, also expressed in full years.

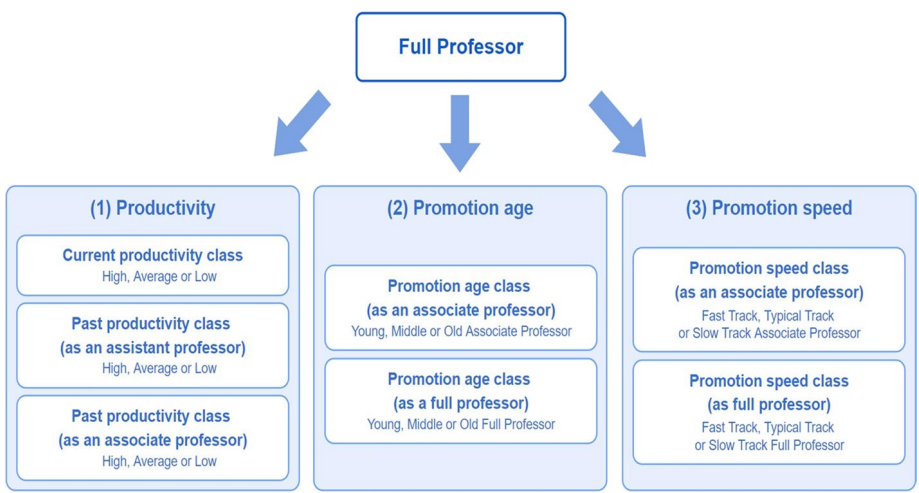

**Fig. 2** Classification scheme used for full professors: productivity, promotion age, and promotion speed classes

At each stage of their careers, the full professors were more productive or less productive. They changed their productivity classes in relation to their colleagues in the same discipline and remained at the same stage of their academic career and in the same academic position. Our study compared "apples with apples" rather than "apples with oranges" (Nygaard et al., 2022). The scientists were consistently compared at the same stage of their careers within the same discipline.

What would not be possible without using raw Scopus (or WoS) metadata in our case? (1) To massively define *disciplines*: we examined all lifetime publications to determine the modal discipline of every full professor. (2) To massively measure *prestige-normalized productivity*: all publications in the lifetime publication histories of all full professors were linked to the journal prestige expressed in the Scopus journal's percentile rank, and 4-year productivity was calculated accordingly. (3) To link *every article* to the three stages of the academic careers of all full professors: only Scopus (or WoS) had all articles by all full professors during their lifetimes. (4) To establish *academic age* for all full professors: the date of the first publication, regardless of type, was necessary in regression models.

### Limitations

The present study has several limitations related to the data and methodology. First, our sample included all Polish scientists who were internationally visible through their research in Scopus from 2009 to 2018 and were employed in Polish higher education system in November 2017; consequently, non-publishing (and non-publishing internationally) scientists were not included. However, the percentage of scientists in STEMM disciplines who published internationally was high; moreover, it increased over time, and it was much higher than in non-STEMM disciplines (Kwiek, 2020).

Second, this research combined (near perfect) administrative and biographical data collected from a national registry of scientists with (much less perfect) bibliometric data at the individual level. Therefore, we combined data on "real individuals" with national identification numbers with metadata on publications by individual Scopus Author IDs rather than "real scientists." Our Observatory of Polish Science was constructed through a deterministic and probabilistic record linkage between two original data sets that differed in nature. For the past two decades, it has been widely debated to what extent bibliometric data are biased linguistically, geographically, and disciplinarily (Boekhout et al., 2021). However, sources other than raw Scopus (or the raw Web of Science Core Collection) datasets could not be used to construct full publication histories of all scientists within a whole national science system. No other source of publication metadata has been available about Polish scientists from the past 50 years. Finally, our study shows a "success bias": its sample includes only full professors i.e. those who got to the top of academic hierarchies.

## Results

### Mobility between productivity classes from a lifetime career perspective

In this subsection, we examine the persistence of productivity classes of full professors from a lifetime career perspective: Have current top-performing full professors always been top-performing? And have current low-performing full professors always been low performing?

Figure 3 shows the lifetime career trajectories of 2326 full professors in 14 STEMM disciplines combined (total). Their productivity was classified as top, middle, or bottom (20%, 60%, or 20%, respectively) in three periods: between becoming assistant professors and becoming associate professors (left column); between becoming associate professors and becoming full professors (middle column); and after becoming full professors (right column). Our focus was on the mobility of top productivity classes and bottom productivity classes in the three stages of an academic career. The results are shown in Sankey diagrams.

The majority of highly productive scientists (top) remained highly productive compared with their peers in the same discipline and within the same academic position, which is shown in thick left-to-right horizontal flows (as shown in Fig. 3). More than half of the highly productive scientists moved from the top class to the top class in the first (52.6%) and second stages of their academic careers (50.8%). Only about 2.3% moved to the low-productivity class in the first period, and only about 5% moved to the low-productivity class in the second period. These exceptional cases of top-to-bottom mobility in productivity classes are shown as thin descending flows from the top classes to the bottom classes (Fig. 3). The mobility from the bottom productivity classes to the top productivity classes in the first and the second periods was limited. In Fig. 3, upward mobility is shown as thin ascending flows from the bottom classes to the top classes: 8.5% and 2.9%, respectively. Extreme mobility between productivity classes (top-to-bottom and bottom-to-top) was characteristic of only 100 scientists of 2326.

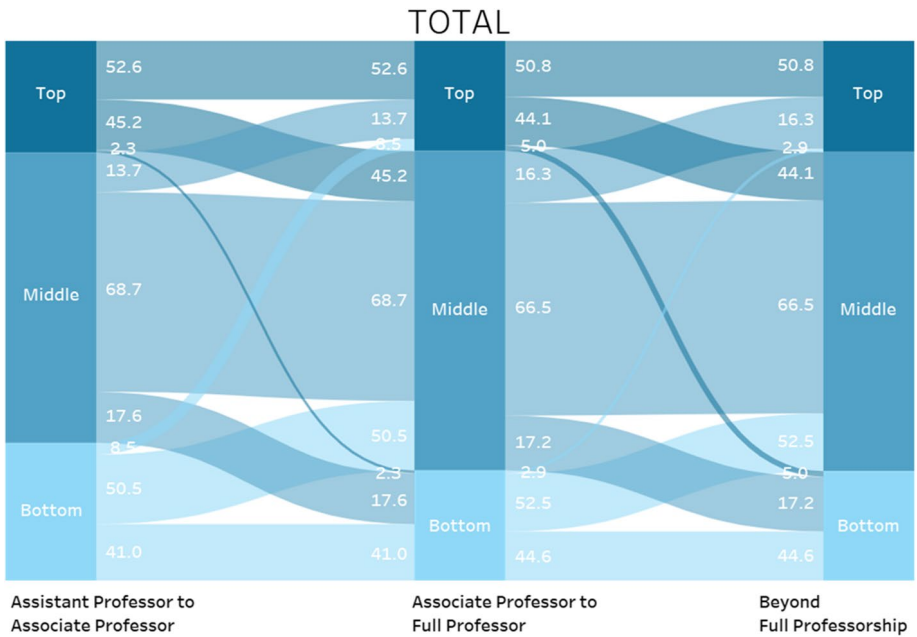

**Fig. 3** Sankey diagram of retrospectively constructed mobility between productivity classes in the three stages of an academic career. All STEMM disciplines (total) are combined, and only current full professors are shown. Top (upper 20%), middle (middle 60%), and bottom (lower 20%) productivity classes are shown in percentages of 100% (or rounded) in each of the three classes. The bottom class in the left column is larger than 20%, and the middle class is smaller than 60%; the cutting-off points did not permit a different division into classes. $N=2326$

The Sankey diagrams also show the ongoing mobility between middle-performing classes (Middle) and top-performing classes (top). Although the majority of professors assigned to the middle-performing class remained in the same class, some moved up, and some moved down. The data on possible combinations of mobility in this case are shown in Table 3: the first panel shows the data on mobility from assistant professors to associate professors, the second panel shows mobility from associate professors to full professors, and the third panel describes the subsample used (all special cases can be identified at an individual level, and further discussed).

Mobility between productivity classes differed substantially between disciplines. We examined in detail the disciplines with the largest number of full professors (i.e., MED) and a discipline in which the patterns of top-to-top and bottom-to-bottom mobility were the most stable from a comparative cross-disciplinary perspective (i.e., MATH). MATH has been frequently studied because of its unique features, such as a low collaboration rate and a low share of female scientists (e.g., Mihaljević-Brandt et al., 2016).

The MED case (Fig. 4) presented a clear pattern of productivity class mobility: its top-to-top and bottom-to-bottom mobility was high, and its top-to-bottom and bottom-to-top mobility was limited over entire academic careers. More than half of highly productive assistant professors (top) became highly productive associate professors (top); and more than half of low-productive assistant professors (bottom) became low-productive associate professors (bottom) (55.1% and 50.6%, respectively; see thick flows in Fig. 4). The mobility pattern was similar for the two stages of academic careers. The majority of highly productive associate professors (top) became highly productive full professors (top), and almost half of the low-productive associate professors (bottom) became low-productive full professors (bottom; 50.6% and 46.1%, respectively). Extreme productivity class transitions (top-to-bottom and bottom-to-top) were rare, which is shown by very thin flows linking top and bottom productivity classes in both periods of their academic careers. Extreme transitions were experienced by 3.8% (downward) and 3.9% (upward) of assistant professors and by 5.2% (downward) and 1.3% (upward) of associate professors.

In the MATH case (Fig. 5), the persistence of highly productive assistants and associate professors was very high. Two-thirds of scientists in the top productivity classes remained in these classes: 69% of highly productive assistant professors continued to be highly productive associate professors, and 65.5% of highly productive associate professors continued to be highly productive full professors. The likelihood that low-productive associate professors would enter the class of highly productive full professors was slim (3.4%).

Overall, cross-disciplinary differences were substantial. An aggregated picture holding for all disciplines combined hides behind it much more nuanced discipline-specific pictures. Disciplines were characterized by different intensities of upward and downward mobility (Fig. 6). In some disciplines, no highly productive assistant professor dropped to the bottom productivity class (e.g., CHEM chemistry, CHEMENG chemical engineering, COMP computer science, EARTH earth and planetary sciences, ENER energy, MATER

Table 3 Mobility between productivity classes in the three stages of academic careers

| Transition from source academic position | Transition from productivity class | Transition to target academic position | Transition to productivity class | Number of scientists in transition | Number of scientists in each productivity class | % |
|---|---|---|---|---|---|---|
| Assistant professor | Bottom | Associate professor | Bottom | 245 | 598 | 41.0 |
| Assistant professor | Bottom | Associate professor | Middle | 302 | 598 | 50.5 |
| Assistant professor | Bottom | Associate professor | Top | 51 | 598 | 8.5 |
| Assistant professor | Middle | Associate professor | Bottom | 222 | 1260 | 17.6 |
| Assistant professor | Middle | Associate professor | Middle | 866 | 1260 | 68.7 |
| Assistant professor | Middle | Associate professor | Top | 172 | 1260 | 13.7 |
| Assistant professor | Top | Associate professor | Bottom | 11 | 485 | 2.3 |
| Assistant professor | Top | Associate professor | Middle | 219 | 485 | 45.2 |
| Assistant professor | Top | Associate professor | Top | 255 | 485 | 52.6 |
| Associate professor | Bottom | Full professor | Bottom | 213 | 478 | 44.6 |
| Associate professor | Bottom | Full professor | Middle | 251 | 478 | 52.5 |
| Associate professor | Bottom | Full professor | Top | 14 | 478 | 2.9 |
| Associate professor | Middle | Full professor | Bottom | 238 | 1387 | 17.2 |
| Associate professor | Middle | Full professor | Middle | 923 | 1387 | 66.5 |
| Associate professor | Middle | Full professor | Top | 226 | 1387 | 16.3 |
| Associate professor | Top | Full professor | Bottom | 24 | 478 | 5.0 |
| Associate professor | Top | Full professor | Middle | 211 | 478 | 44.1 |
| Associate professor | Top | Full professor | Top | 243 | 478 | 50.8 |
| Full professor | Bottom | | | 475 | 475 | 100 |
| Full professor | Middle | | | 1385 | 1385 | 100 |
| Full professor | Top | | | 483 | 483 | 100 |

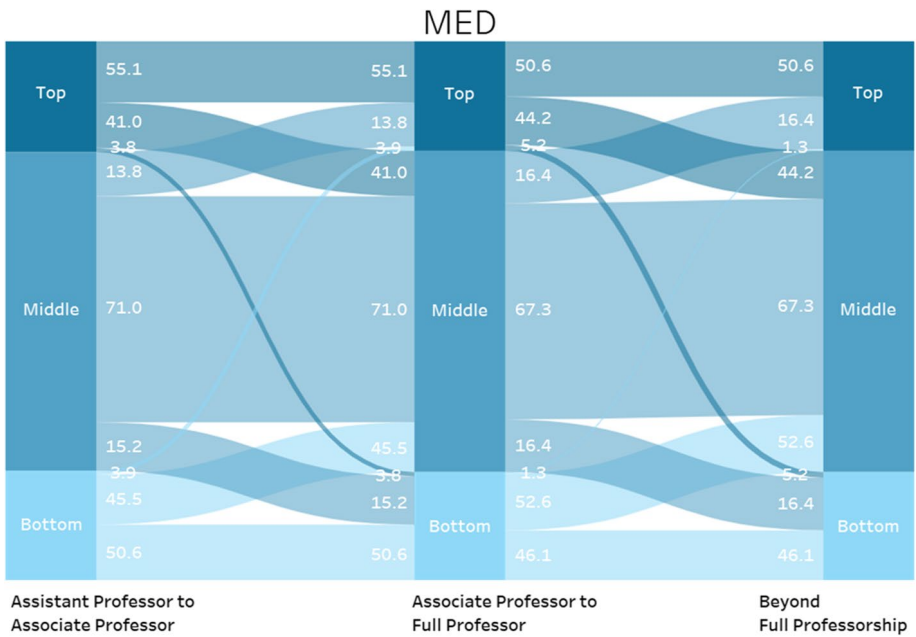

**Fig. 4** Sankey diagram of retrospectively constructed mobility between productivity classes in the three stages of academic careers. MED and current full professors only. $N=379$

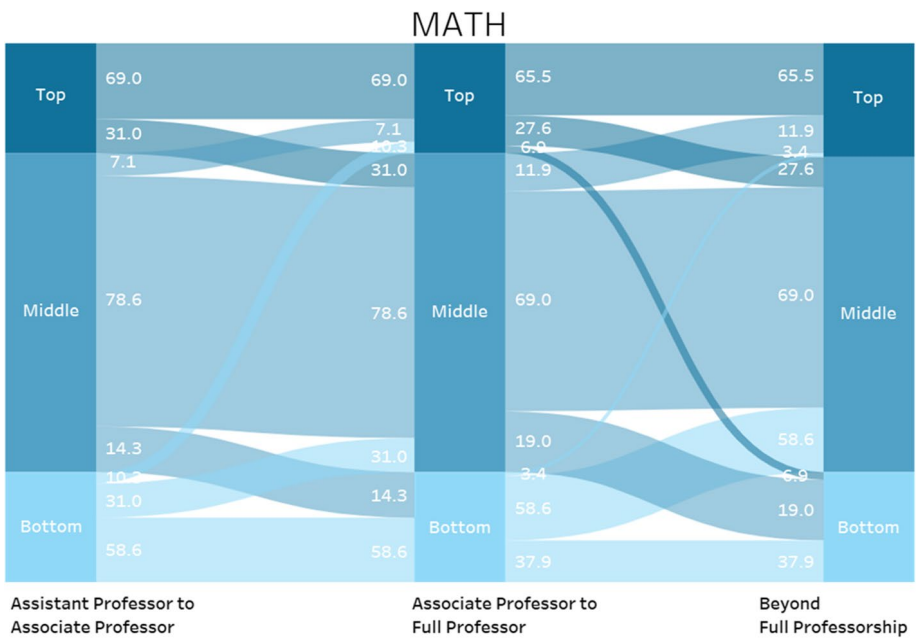

**Fig. 5** Sankey diagram of retrospectively constructed mobility between productivity classes in the three stages of academic careers. MATH and current full professors only. $N=142$

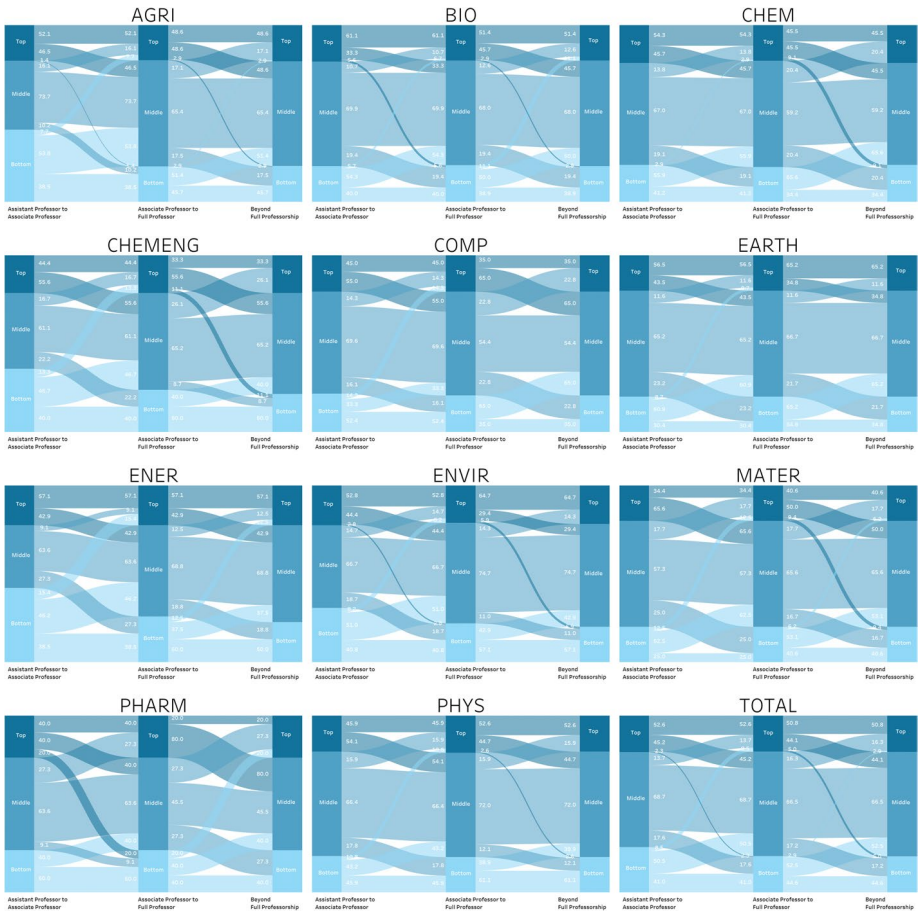

**Fig. 6** Overview: Sankey diagrams of retrospectively constructed mobility between productivity classes in the three stages of academic careers. Eleven STEMM disciplines and all disciplines combined (total), current full professors only materials science, and MATH mathematics). Upward mobility from a bottom class to a top class was nonexistent for associate professors in CHEMENG chemical engineering, EARTH earth and planetary sciences, ENVIR environment, and PHYS physics and astronomy. In other disciplines, no highly productive assistant professor and no highly productive associate professor dropped to the bottom productivity class, and upward mobility to a top class was nonexistent for associate professors (e.g., computer science [COMP] and earth and planetary sciences [EARTH]). In other disciplines, while no top-to-bottom mobility in productivity classes was observed, bottom-to-top mobility was notable (e.g., energy [ENER] and physics and astronomy [PHYS]). Finally, the biggest variability, as expected, was noted for the smallest disciplines, as the case of PHARM pharmacology, toxicology, and pharmaceutics clearly shows. Moreover, the results showed variations by gender within disciplines in which higher proportions of women than men remained in the top productivity classes, as shown in Tables 4 and 5.

**Table 4** Retrospectively constructed selected mobility between productivity classes (top to top, bottom to bottom) in the three stages of academic careers. Current full professors only, by discipline and all STEMM disciplines combined (total). The 20/60/20 division: top class (upper 20%), middle class (middle 60%), and bottom class (lower 20%) in percentages, 100% (or rounded) in columns. $N = 2326$

| Discipline | Mobility: three academic career stages (assistant professor → associate professor → full professor) | | | | | | | | | | | |
|---|---|---|---|---|---|---|---|---|---|---|---|---|
| | Mobility 1 (top to top): from assistant professor top class to associate professor top class | | | Mobility 2 (top to top): from associate professor top class to full professor top class | | | Mobility 3 (bottom to bottom): from assistant professor bottom class to associate professor bottom class | | | Mobility 4 (bottom to bottom): from associate professor bottom class to full professor bottom class | | |
| | Male | Female | Total | Male | Female | Total | Male | Female | Total | Male | Female | Total |
| AGRI | 52.2 | 52.1 | 52.1 | 26.9 | 61.4 | 48.6 | 33.3 | 41.1 | 38.5 | 56.2 | 42.6 | 45.7 |
| BIO | 46.2 | 69.6 | 61.1 | 37.5 | 63.2 | 51.4 | 46.2 | 36.4 | 40.0 | 40.0 | 38.1 | 38.9 |
| CHEM | 50.0 | 56.5 | 54.3 | 27.3 | 54.5 | 45.5 | - | 43.8 | 41.2 | 75.0 | 28.6 | 34.4 |
| CHEMENG | 33.3 | 50.0 | 44.4 | 33.3 | 33.3 | 33.3 | - | 42.9 | 40.0 | 100.0 | 55.6 | 60.0 |
| COMP | - | 45.0 | 45.0 | 33.3 | 35.3 | 35.0 | 50.0 | 53.3 | 52.4 | 16.7 | 42.9 | 35.0 |
| EARTH | 66.7 | 55.0 | 56.5 | 33.3 | 70.0 | 65.2 | - | 33.3 | 30.4 | 33.3 | 35.0 | 34.8 |
| ENER | 100.0 | 50.0 | 57.1 | 100 | 50.0 | 57.1 | 75.0 | 22.2 | 38.5 | 66.7 | 40.0 | 50.0 |
| ENG | 100.0 | 48.3 | 50.8 | 40 | 49.1 | 48.4 | 50.0 | 34.2 | 35.0 | 33.3 | 44.1 | 43.5 |
| ENVIR | 41.7 | 58.3 | 52.8 | 62.5 | 65.4 | 64.7 | 39.1 | 42.3 | 40.8 | 46.2 | 63.6 | 57.1 |
| MATER | 62.5 | 25.0 | 34.4 | 37.5 | 43.8 | 40.6 | - | 32.0 | 25.0 | - | 40.6 | 40.6 |
| MATH | 100.0 | 67.9 | 69.0 | 50 | 66.7 | 65.5 | 75.0 | 56.0 | 58.6 | - | 42.3 | 37.9 |
| MED | 56.7 | 54.2 | 55.1 | 37.5 | 60.0 | 50.6 | 54.5 | 47.7 | 50.6 | 55.2 | 40.4 | 46.1 |
| PHARM | 50.0 | 33.3 | 40.0 | 25 | - | 20.0 | 75.0 | - | 60.0 | 33.3 | 50.0 | 40.0 |
| PHYS | 100.0 | 44.4 | 45.9 | 33.3 | 54.3 | 52.6 | - | 47.2 | 45.9 | - | 62.9 | 61.1 |
| **Total** | **54.9** | **51.9** | **52.6** | **35.8** | **56.7** | **50.8** | **41.2** | **40.9** | **41.0** | **47.0** | **43.9** | **44.6** |

*Note.* - = no observation (no full professor in this class)

**Table 5** Retrospectively constructed selected mobility between productivity classes (top to top, bottom to bottom) in the two stages of academic careers. Current full professors only, by discipline and all STEMM disciplines combined (total). The 20/60/20 division: top class (upper 20%), middle class (middle 60%), and bottom class (lower 20%) in percentages, 100% (or rounded) in columns. $N = 2326$

| Discipline | Mobility: two academic career stages (assistant professor → full professor) | | | | | |
|---|---|---|---|---|---|---|
| | Mobility 5 (bottom to bottom): from assistant professor bottom class to full professor bottom class | | | Mobility 6 (top to top): from assistant professor top class to full professor top class | | |
| | Male | Female | Total | Male | Female | Total |
| AGRI | 35.4 | 31.6 | 32.9 | 39.1 | 50.0 | 46.5 |
| BIO | 30.8 | 31.8 | 31.4 | 23.1 | 52.2 | 41.7 |
| CHEM | 50.0 | 25.0 | 26.5 | 41.7 | 52.2 | 48.6 |
| CHEMENG | 100.0 | 21.4 | 26.7 | 33.3 | 16.7 | 22.2 |
| COMP | - | 13.3 | 9.5 | - | 30.0 | 30.0 |
| EARTH | 50.0 | 33.3 | 34.8 | - | 60.0 | 52.2 |
| ENER | 50.0 | 55.6 | 53.8 | 100.0 | 50.0 | 57.1 |
| ENG | 25.0 | 30.3 | 30.0 | 33.3 | 45.0 | 44.4 |
| ENVIR | 26.1 | 50.0 | 38.8 | 58.3 | 50.0 | 52.8 |
| MATER | - | 32.0 | 25.0 | 62.5 | 33.3 | 40.6 |
| MATH | - | 40.0 | 34.5 | 100 | 46.4 | 48.3 |
| MED | 51.5 | 40.9 | 45.5 | 46.7 | 62.5 | 56.4 |
| PHARM | 50.0 | - | 40.0 | - | 33.3 | 20.0 |
| PHYS | - | 47.2 | 45.9 | 100.0 | 44.4 | 45.9 |
| **Total** | **34.0** | **34.2** | **34.1** | **42.5** | **48.1** | **46.8** |

*Note.* - = no observation (no full professor in this class).

Figure 6 shows transitions in all disciplines not described above. The stability of top-performing classes was high and ranged from 34.4–69.0% for assistant professors who became associate professors and 20–65.5% for associate professors who became full professors. The share exceeded 50% in most disciplines in the first case and in half of the disciplines in the second case. Further details on mobility are shown in Table 4.

We also conducted a comparison of productivity classes in the first and last stages of the academic career (Table 5): assistant professor and full professor. Almost half of the current highly productive full professors had been highly productive assistant professors 20–40 years earlier (46.8%). However, the results showed an interesting gender disparity: the percentage of female scientists who continued to be highly productive throughout their careers was considerably higher than the percentage of male scientists who continued to be highly productive throughout their careers (48.1% vs. 42.5%) (Table 5). Cross-disciplinary and gender differences were substantial: for instance, all (100%) highly productive male full professors were highly productive assistant professors 20–40 years earlier in the two most male-dominated disciplines, MATH and PHYS (compared with females at 46.4% and 44.4%, respectively). The principle "once highly productive, forever highly productive" held for all cases of Polish male mathematicians, physicists, and astronomers (current full professors).[1]

## Logistic regression models

This subsection presents the odds ratio estimates of belonging to top productivity classes for current full professors and, retrospectively, for current full professors at earlier stages of their academic careers (in the same disciplines) ($N=2326$). The individual-level variables included gender, biological age, academic age (the number of years since the first publication, see Kwiek and Roszka, 2022b), and the biological ages at which the doctorate, habilitation (or postdoctoral degree), and full professorship were awarded. Most importantly in the context of the two-dimensional analyses presented in the section "Mobility between productivity classes from a lifetime career perspective," the individual-level variables also included classifications from our general classificatory scheme (Fig. 2): membership in current and past productivity classes, promotion-age classes, and promotion-speed classes (with the 20/60/20 divisions in each case). The only organization-level variable was the research intensity of the employing institution (IDUB vs. other institutions); other variables were tested (e.g., research budget, total budget, total number of scientists) but proved significantly correlated with the IDUB variable.

Crucially, the results of the logistic regression (Table 6) powerfully augment the result yielded by the descriptive statistics: being among the most productive full professors is dependent on having been in analogous groups of highly productive scientists at earlier stages of one's academic career. Thus, belonging to the class of highly productive assistant professors increased the probability of becoming a highly productive full professor by, on average, from two up to almost four times (Exp(B)=2.8; 95% confidence interval 2.1–3.6), while belonging to the class of highly productive associate professors increased the probability of success by, on average, from almost four to almost six times (Exp(B)=4.61; 95% confidence interval 3.6–6). The only significant predictor indirectly related to age was belonging to the youngest 20% of full professors in terms of promotion age. Membership in this class increased the probability of success by, on average, almost twice (Exp(B)=1.942; see the variables Top_assistant_prof_class, Top_associate_prof_class, and Young_full_prof_class).

Similarly, among current full professors when they were associate professors (Model 2), belonging to the class of highly productive assistant professors increased the probability of becoming a highly productive associate professor by, on average, from almost five to more than nine times (Exp(B)=6.667; 95% confidence interval 4.7–9.4). The important determinants of membership in the top 20% of productive scientists were related to age, both biological and academic. Biological age had a negative effect, and it had a significantly stronger negative effect on associate professors than on assistant professors. An increase in biological age by one year reduced the probability of entering the class of highly productive assistant professors by 20–25%. Among associate professors, this 1-year increase reduced the likelihood by up to one-third – one-fourth. Among assistant professors, a 1-year increase in academic age (and thus publication experience or the number of years since first publication) resulted in an average increase of 10–15% in the probability of success, while, among associate professors, the average increase was only 0.2–4.1%.

---

[1] Z-test statistics for the three stages of academic careers show that for both transitions between assistant professors, associate professors and full professors, as well as for direct transitions from assistant professorship to full professorship, in most domains no differences in percentages were observed for men and women. Exceptions included top associate professor to top full professor transitions, where a significant difference was observed for AGRI ($z=-2.786$, $p$-value=0.008, Cohen's $d=0.333$) and for all domains ($z=-3.955$, $p$-value<0.001, Cohen's $d=0.182$). A significant difference was observed for ENVIR for the transition from bottom assistant professors to bottom full professors ($z=-2.384$, $p$-value=0.023, Cohen's $d=0.341$). In all above cases, a significantly higher percentage was observed for women. The effect size of the differences is rather moderate.

**Table 6** Logistic regression statistics: odds ratio estimates of belonging the classes of highly productive full professors, associate professors, and assistant professors (the upper 20%, separately for each discipline; current full professors only, $N=2326$)

| Model | Model 1: full professors $R^2=0.254$ | | | | Model 2: associate professors $R^2=0.582$ | | | | Model 3: assistant professors $R^2=0.355$ | | | |
| --- | --- | --- | --- | --- | --- | --- | --- | --- | --- | --- | --- | --- |
| | Exp(B) | 95% C.I. for EXP(B) | | Sig | Exp(B) | 95% C.I. for EXP(B) | | Sig | Exp(B) | 95% C.I. for EXP(B) | | Sig |
| | | Lower | Upper | | | Lower | Upper | | | Lower | Upper | |
| Male | | | | | 1.426 | 1.03 | 1.974 | 0.033 | | | | |
| Research intensive | | | | | | | | | | | | |
| Biological age | | | | | 0.694 | 0.665 | 0.724 | <0.001 | 0.774 | 0.753 | 0.796 | <0.001 |
| Academic age | | | | | 1.021 | 1.002 | 1.041 | 0.028 | 1.122 | 1.098 | 1.148 | <0.001 |
| Assistant_age | | | | | 0.942 | 0.892 | 0.995 | 0.032 | 1.207 | 1.143 | 1.273 | <0.001 |
| Associate_age | | | | | 1.475 | 1.404 | 1.549 | <0.001 | - | - | - | - |
| Full_age | | | | | - | - | - | - | - | - | - | - |
| Top_assistant | 2.793 | 2.14 | 3.646 | <0.001 | 6.667 | 4.72 | 9.416 | <0.001 | - | - | - | - |
| Top_associate | 4.61 | 3.558 | 5.974 | <0.001 | - | | | | - | - | - | - |
| Young_assistant | | | | | | | | | 1.739 | 1.232 | 2.455 | 0.002 |
| Young_associate | | | | | - | | | | - | - | - | - |
| Young_full | 1.942 | 1.503 | 2.509 | <0.001 | - | | | | - | - | - | - |
| Fast_associate | | | | | - | | | | - | - | - | - |
| Fast_full | | | | | - | | | | - | - | - | - |
| Constant | 0.1 | | | <0.001 | 46.17 | | | <0.001 | 128.62 | | | <0.001 |

Another age-related variable that significantly affected the probability of success was the promotion age of assistant professors. Among associate professors, an increase in doctoral promotion age (variable: Assistant_prof_promotion_age) had a negative effect, decreasing the probability of success by an average of 5.8% (with 95% confidence interval 0.5–10.8%), while, among assistant professors, the direction of change was positive and high; a 1-year increase in doctoral promotion age increased the probability of success by an average of 20.7% (14–27%). The age at promotion to a postdoctoral degree (variable: Associate_prof_promotion_age) significantly and strongly influenced the probability of success; a 1-year increase in the age at promotion increased the likelihood of entering the group of the 20% most productive associate professors by about half (on average, 47.5%; 40–55%). This variable could not be included in the model for assistant professors because they had not yet been promoted to this stage, having not yet earned their postdoctoral degrees. However, one variable (indirectly) related to age that was important to the likelihood of being among the 20% of most productive assistant professors was being among the 20% of the youngest scientists promoted to doctoral degrees. Membership in this group increased the probability of success by an average of 73.9 (although the confidence interval in this case was quite wide: 23.2–145.5%). Gender had a significant impact only among associate professors. Being male increased the probability of success by an average of 42.6%, but the range of the confidence interval (3%–97%) suggests that the significance of this predictor should be interpreted with caution (the role of gender differences, see Kwiek and Roszka, 2021a, b; Kwiek and Roszka, 2022a).

In summary, for current full professors, the most powerful predictors of belonging to the class of highly productive scientists are having belonged to that class while working as assistant professors and as associate professors; a third powerful predictor is membership in the class of full professors promoted early in their careers. Retrospectively, for current full professors in their past as associate professors, the single most powerful predictor is having belonged to the class of highly productive assistant professors; other predictors include belonging to the class of associate professors promoted early (Exp(B) = 1.475) and, possibly, being a male (Exp(B) = 1.426). Finally, also retrospectively for assistant professors, the single most powerful predictor of belonging to the class of highly productive assistant professors is belonging to the class of assistant professors promoted early (Exp(B) = 1.739).

## Discussion and conclusions

Highly productive scientists have often been examined as a special academic class: as "eminent" and "highly prolific" scientists and as "stars," "top scientists," and "top performers" (Fox & Nikivincze, 2021; Agrawal et al., 2017; Kwiek, 2016; Cortés et al., 2016; Abramo et al., 2009). They are "motivated by an inner drive to do science and by a sheer love of the work" (Cole & Cole, 1973: 62), and, while some scientists are particularly good at doing science, "some are not just good but superb" (Stephan & Levin, 1992: 13). In that vein, some full professors in our sample were simply superb at doing science from the moment they entered academia through their late-career stages. About half the highly productive full professors had always been highly productive, regardless of the trajectories of their personal lives or their external circumstances (e.g., the post-communist transition period in the Polish economy, which severely affected the academic sector). Highly productive full professors in their 60s were also highly productive when they were assistant and associate professors in their 30s, 40s, and 50s.

The message of our regression analysis is simple: past productivity classes (i.e., publication history) powerfully determine current productivity classes, with much smaller roles played by the other predictors. Our regression models strongly support the results of our two-dimensional analyses, according to which scientists who have once been highly productive tend to remain highly productive and those who have once had poor productivity have little chance of moving to the high productivity classes (shown as thin upward flows between the bottom and top productivity classes across all disciplines) (Fig. 3).

There are only two powerful predictors of high productivity among full professors: membership in the class of highly productive assistant professors and membership in the class of highly productive associate professors, which increase the odds by, on average, almost three and five times, respectively (by 179% and 361%). The most powerful predictor of becoming a highly productive associate professor (in the sample of current full professors) was being a highly productive assistant professor as shown by the staggering increase in odds: almost seven times (or by 570%). For highly productive assistant professors, the most powerful predictor was obtaining a PhD early in their careers. Additionally, our results support previous findings that full professors appointed early tend to be more productive than full professors appointed later in their careers (Abramo et al., 2016). Membership in the class of young full professors increased the odds of belonging to the class of highly productive full professors by an average of 94.2%. Neither gender nor age (biological or academic) emerged as a predictor of membership in the class of highly productive full professors. The results did not directly support the claim that the productivity of top- and medium-performing scientists increases or remains stable with age (Costas et al., 2010: 1578), as our study focused on changing productivity classes rather than on evolving productivity over time.

The results of our study revealed an unexpectedly high level of immobility in the system. Membership in the productivity class during assistant professorships and associate professorships, to a large extent, determined membership in the productivity class during full professorship and beyond. Does the "once highly productive, forever highly productive" principle hold across all STEMM disciplines? The results of this study indicate the affirmative. About half of the current full professors belonged to the same productivity class throughout their academic careers. They had remained for decades in the bottom or top productivity classes in relation to their peers and within their specific disciplines. About half of the current full professors had changed their productivity class membership by only one class in a tripartite division into top, middle, or bottom classes, with some discipline and gender differentiation. Cross-disciplinary and gender differences were substantial: for instance, all highly productive male full professors (100%) were highly productive assistant professors 20–40 years earlier in the two most male-dominated disciplines, MATH and PHYS. So the principle held for all cases of Polish male mathematicians, physicists, and astronomers.

More than half of the highly productive assistant professors became, on average, highly productive associate professors in relation to their peers in a similar period, the same academic position, and the same discipline. More than half of the highly productive associate professors became, on average, highly productive full professors (52.6% and 50.8%, respectively). Moreover, a study of direct start-to-end mobility shows that, on average, almost half of the highly productive assistant professors became highly productive full professors. They did not change their productivity class membership to a lower class throughout their academic careers (46.8%), with a large differentiation among disciplines. Similar processes of transition in productivity class membership included low-productive scientists.

The most radical changes in productivity class membership, that is, transitions from the very top to the very bottom of productivity, occurred at a marginal level; upward bottom-to-top transfers occurred on a similar small scale. In our sample, the 2326 full professors in the last four decades included 35 scientists who had radically changed their productivity classes downward and 65 who had moved upward (so, in total, only 4.3% of current full professors). Above-average mobility was observed in the disciplines of BIO, MATH, and PHYS, while the least mobility was observed in PHARM.

Perhaps the most interesting question is why the pattern of "once highly productive, forever highly productive" is so pervasive in Polish higher education. Among several possible explanations, one follows the lines of two traditional theories of productivity, sacred spark theory and cumulative advantage theory. The former holds that there is a small group of scientists who will always be superb in their achievements, as they have a spark that others lack, being inherently highly motivated, well organized, creative, and skillful. The latter theory identifies a group of scientists who, with or without that spark, keep accumulating advantages from the very beginning of their careers. Their advantages come from their socialization to internationalized work environments, specific work cultures, and work habits available mostly in elite institutions or departments; from doctoral advisors who were role models; and from resources available through research funding, including long-term international fellowships. The cumulative advantage theory explains high productivity by a set of reinforcing factors that, combined, continually push academic careers forward (with ever better access to resources of all kinds: research time, infrastructure, funding, international networks, publications in prestigious journals, externally funded doctoral and postdoctoral researchers, etc.).

Another useful theoretical line of explanation is the credibility cycle in academic careers (Latour & Woolgar, 1986: 200–208), in which prestigious papers are converted into recognition that leads to successful individual grant applications that are converted into new equipment, data, software, arguments, and articles. Perhaps the credibility cycle is faster for scientists affected by this mechanism at an early stage in their careers: once funded, with excellent publications, they have better chances to be funded again and to be promoted sooner to higher ranks, reflecting the idea that each element of the credibility cycle in academic careers "is but one part of an endless cycle of investment and conversion" (Latour & Woolgar, 1986: 200). Advantages already gained lead more quickly to future advantages, as in any positional competition having the nature of a zero-sum game: "what winners win, losers lose" (Hirsch, 1976: 52). The above theoretical mechanisms have more powerful effects in resource-poor systems such as Poland's, in which, historically, funding could be won or lost by a small margin due to the scarcity of public funding.

The patterns of mobility between productivity classes over the course of an entire academic career in national academic science systems may have far-reaching implications for science policies, especially regarding hiring and promotion. Hiring and tenure to both low-productivity and high-productivity scientists may have long-standing consequences for institutions and the national system in terms of the average productivity level. Research careers are usually long. After entering the system and achieving job stability, scientists in Poland (where attrition is very low) and elsewhere usually remain in the system for years, if not decades (see especially Abramo et al., 2017 discussing star scientists and unproductive scientists in Italy). The scientists included in this study, all of whom are currently full professors in the 14 STEMM disciplines present in the Scopus bibliometric database, have remained in the system for 20–40 years. Individual hiring and promotion decisions made at the departmental level thus have long-lasting implications for productivity at the national level, spanning two to four decades.

Our results may also imply the need to cultivate productivity, especially among young academics: entering productivity elites early on substantially increases the chances of belonging to productivity elites in later career stages. The importance of cultivating productivity goes beyond research-intensive universities and pertains to the whole higher education sector. Understanding persistent inequality in productivity matters especially in resource-poor systems in which research funding is highly competitive. Scientists in STEMM disciplines tend to be powerfully locked-in early on in their careers in their productivity classess and the chances of changing them radically from a long-term longitudinal perspective—becoming much more productive compared with their peers—are slim. It would be interesting to see whether similar mechanisms operate within social sciences and humanities; however, the character of the Scopus data (limited coverage for social sciences and humanities in the Polish case) does not allow us to go beyond STEMM disciplines in our research.

The results of our study indicate the opportunities provided by structured Big Data (in this case, the Scopus raw dataset). We examined all current Polish full professors in STEMM, but the data we used were collected from two large datasets. One was the Observatory of Polish Science, which included full biographical and administrative data on almost 100,000 Polish scientists and their 380,000 publications in Scopus from 2009–2018. The second dataset comprised Scopus metadata on almost a million (935,167) Polish publications in the past 50 years. The merger of several datasets made it possible to create not only current productivity classes to which all professors were allocated – but also retrospective productivity classes. Importantly, every full professor was compared in terms of research productivity as an assistant and associate professor with their exact peers (current full professors) when they were at the same earlier stages of academic careers in the same discipline. We retrospectively examined their academic careers as extensively as necessary to compare "apples with apples" rather than "apples with oranges" at all three stages of their academic careers.

Finally and more generally, structured Big Data offer fundamentally new opportunities to examine the academic profession, both nationally, cross-nationally, and globally. The Big Data collected and stored by others (e.g., governments and corporations) for other than academic purposes can be analyzed by students of the academic profession as a new, complementary data source to complement traditional sources, such as academic surveys and interviews. This could provide a better balance between small-scale (low-N) and large-scale (big-N) studies, with a fertilizing effect on the field (for an overview of the field, see Carvalho, 2017). The key word is *complementarity*: new data sources complement, rather than replace, traditional sources.

The new data must be repurposed (Salganik, 2018), and they come with their own limitations and biases, but the amount of data available and their longitudinal character (enabling the analysis of changes in academic careers over time) offer great promise. From datasets that are vast in both size and complexity, we can extract only the useful current and past information about academics and their output. We can examine huge amounts of data to discover patterns that would otherwise be imperceptible, looking at outliers, deviations, and special cases and performing analyses based on unprecedented numbers of observations. While Big Data dramatically deepen our insight into society generally (Selwyn, 2019), specific parts of structured, curated, and reliable Big Data (such as commercial bibliometric datasets) can radically sharpen our insights into the academic profession, allowing it to be examined with the use of new temporal (time), topical (themes), geographical (places), and network (connections) analyses (see Börner, 2010: 62–63). The various dimensions of academic work can be studied with ever more precision and a remarkable level of detail.

The use of curated, large-scale data sources allows to study the academic profession over years, across countries (institutions, cities), across academic disciplines, at different levels of granularity and in terms of research teams and individuals, male and female scientists, and junior and senior scientists. The small observation numbers yielded by traditional surveys of the academic profession limit the analytical power of the datasets and weaken the ability to draw policy implications from the research. Small-scale studies are useful and theoretically inspiring but, in the current world, they may not convince policy makers and grant-making agencies. Several factors increase the pressure to study the academic profession using Big Data: first, the increasing availability of digital data on scholarly inputs and outputs at an individual level (funding, publications, collaboration, mobility); second, the growing availability of computing power to analyze the data; finally, the pressure to provide both the public and the scholarly community with a more quantified, data-based, sound, and convincing understanding of changes in higher education in general and in the academic profession in particular.

**Acknowledgements** Marek Kwiek is grateful for the comments from the hosts and audiences of seminars at the University of Oxford (CGHE, Center for Global Higher Education), Stanford University (METRICS, Meta-Research Innovation Center at Stanford), both in June 2022, and DZHW (German Center for Higher Education Research and Science Studies), Berlin, in January 2023. We gratefully acknowledge the assistance of the International Center for the Studies of Research (ICSR) Lab and Kristy James, Senior Data Scientist. We also want to thank Lukasz Szymula from the CPPS Poznan Team for improving the visualizations. We gratefully acknowledge the support provided by the NDS grant no. NdS/529032/2021/2021.

**Data Availability** We used data from Scopus, a proprietary scientometric database. For legal reasons, data from Scopus received through collaboration with the International Center for the Studies of Research (ICSR) Lab cannot be made openly available.

## Declarations

**Conflict of interest** The authors declare no competing interests.

## References

Abramo, G., D'Angelo, C. A., & Caprasecca, A. (2009). The contribution of star scientists to overall sex differences in research productivity. *Scientometrics, 81*(1), 137–156.
Abramo, G., D'Angelo, C. A., & Di Costa, F. (2011). Research productivity: Are higher academic ranks more productive than lower ones? *Scientometrics, 88*(3), 915–928.

Abramo, G., D'Angelo, C. A., & Murgia, G. (2016). The combined effects of age and seniority on research performance of full professors. *Science and Public Policy, 43*(3), 301–319.

Abramo, G., D'Angelo, C. A., & Soldatenkova, A. (2017). How long do top scientists maintain their stardom? An analysis by region, gender and discipline: Evidence from Italy. *Scientometrics, 110*, 867–877.

Agrawal, A., McHale, J., & Oettl, A. (2017). How stars matter: Recruiting and peer effects in evolutionary biology. *Research Policy, 46*(4), 853–867.

Aguinis, H., & O'Boyle, E. (2014). Star performers in twenty-first century organizations. *Personnel Psychology, 67*(2), 313–350.

Aksnes, D. W., Rørstad, K., Piro, F. N., & Sivertsen, G. (2011). Are female researchers less cited? A large-scale study of Norwegian researchers. *Journal of the American Society for Information Science and Technology, 62*(4), 628–636.

Allison, P. D., & Stewart, J. A. (1974). Productivity differences among scientists: Evidence for accumulative advantage. *American Sociological Review, 39*(4), 596–606.

Boekhout, H., van der Weijden, I., & Waltman, L. (2021). Gender differences in scientific careers: A large-scale bibliometric analysis. https://arxiv.org/abs/2106.12624. Accessed 10 Jan 2023.

Börner, K. (2010). *Atlas of Science*. The MIT Press.

Carvalho, T. (2017). The study of the academic profession—Contributions from and to the sociology of professions. In J. Huisman & M. Tight (Eds.), *Theory and method in higher education research* (pp. 59–76). Emerald.

Cole, J. R., & Cole, S. (1973). *Social stratification in science*. University of Chicago Press.

Cortés, L. M., Mora-Valencia, A., & Perote, J. (2016). The productivity of top researchers: A semi-nonparametric approach. *Scientometrics, 109*(2), 891–915.

Costas, R., van Leeuwen, T. N., & Bordons, M. (2010). Self-citations at the meso and individual levels: Effects of different calculation methods. *Scientometrics, 82*, 517–537.

DiPrete, T. A., & Eirich, G. M. (2006). Cumulative advantage as a mechanism for inequality: A review of theoretical and empirical developments. *Annual Review of Sociology, 32*(1), 271–297.

Fox, M. F. (1983). Publication productivity among scientists: A critical review. *Social Studies of Science, 13*(2), 285–305.

Fox, M. F. (2020). Gender, science, and academic rank: Key issues and approaches. *Quantitative Science Studies, 1*(3), 1001–1006.

Fox, M. F., & Mohapatra, S. (2007). Social-organizational characteristics of work and publication productivity among academic scientists in doctoral-granting departments. *Journal of Higher Education, 78*(5), 542–571.

Fox, M. F., & Nikivincze, I. (2021). Being highly prolific in academic science: Characteristics of individuals and their departments. *Higher Education, 81*, 1237–1255.

GUS. (2022). *Higher education and its finances in 2021*. Main Statistical Office of Poland.

Habicht, I. M., Lutter, M., & Schröder. (2022). Gender differences in the determinants of becoming a professor in Germany. An event history analysis of academic psychologists from 1980 to 2019. *Research Policy, 51*(6), 104506.

Heckman, J. J., & Moktan, S. (2018). *Publishing and promotion in economics. The tyranny of the top five*. NBER (Working Paper 25093).

Hermanowicz, J. (2012). The sociology of academic careers: Problems and prospects. In J. C. Smart & M. B. Paulsen (Eds.), *Higher education: Handbook of theory and research 27* (pp. 207–248)

Hirsch, F. (1976). *Social limits to growth*. Harvard University Press.

Katz, D. A. (1973). Faculty salaries, promotions, and productivity at a large university. *American Economic Review, 63*(3), 469–477.

Kolesnikov, S., Fukumoto, E., & Bozeman, B. (2018). Researchers' risk-smoothing publication strategies: Is productivity the enemy of impact? *Scientometrics, 116*(3), 1995–2017.

Kyvik, S. (1990). Age and scientific productivity: Differences between fields of learning. *Higher Education, 19*, 37–55.

Kwiek, M. (2015). Academic generations and academic work: Patterns of attitudes, behaviors and research productivity of Polish academics after 1989. *Studies in Higher Education*, *40*(8), 1354–1376.

Kwiek, M. (2016). The European research elite: A cross-national study of highly productive academics across 11 European systems. *Higher Education, 71*(3), 379–397.

Kwiek, M. (2018). High research productivity in vertically undifferentiated higher education systems: Who are the top performers? *Scientometrics, 115*(1), 415–462.

Kwiek, M. (2019). Changing European academics. In *A comparative study of social stratification, work patterns and research productivity*. London and New York: Routledge.

Kwiek, M. (2020). Internationalists and locals: International research collaboration in a resource-poor system. *Scientometrics*, *124*, 57–105.
Kwiek, M., & Roszka, W. (2021a). Gender disparities in international research collaboration: A large-scale bibliometric study of 25,000 university professors. *Journal of Economic Surveys*, *35*(5), 1344–1388.
Kwiek, M., Roszka, W. (2021b). Gender-based homophily in research: A large-scale study of man-woman collaboration. *Journal of Informetrics*, *15*(3), 1–38.
Kwiek, M., Roszka, W. (2022a). Are female scientists less inclined to publish alone? The gender solo research gap. *Scientometrics*, *127*, 1697–1735.
Kwiek, M., Roszka, W. (2022b). Academic vs. biological age in research on academic careers: a large-scale study with implications for scientifically developing systems. *Scientometrics*, *127*, 3543–3575.
Latour, B., & Woolgar, S. (1986). *Laboratory Life*. Princeton University Press.
Lerchenmueller, M. J., & Sorenson, O. (2018). The gender gap in early career transitions in the life sciences. *Research Policy, 47*(6), 1007–1017.
Lutter, M., & Schröder, M. (2016). Who becomes a tenured professor, and why? Panel data evidence from German sociology, 1980–2013. *Research Policy, 45*(5), 999–1013.
Madison, G., & Fahlman, P. (2020). Sex differences in the number of scientific publications and citations when attaining the rank of professor in Sweden. *Studies in Higher Education, 46*(12), 2506–2527.
Marginson, S. (2014). University research: The social contribution of university research. In J. C. Shin & U. Teichler (Eds.), *The future of the post-massified university at the crossroads* (pp. 101–118). Springer International Publishing.
Marini, G., & Meschitti, V. (2018). The trench warfare of gender discrimination: Evidence from academic promotions to full professor in Italy. *Scientometrics, 115*(2), 989–1006.
Merton, R. K. (1973). *The sociology of science: Theoretical and empirical investigations*. University of Chicago Press.
Mihaljević-Brandt, H., Santamaría, L., & Tullney, M. (2016). The effect of gender in the publication patterns in mathematics. *PLoS ONE, 11*(10), e0165367.
Nygaard, L. P., Aksnes, D. W., & Piro, F. N. (2022). Identifying gender disparities in research performance: The importance of comparing apples with apples. *Higher Education, 84*, 1127–1142.
Piro, F. N., Rørstad, K., & Aksnes, D. W. (2016). How do prolific professors influence the citation impact of their university departments? *Scientometrics, 107*(3), 941–961.
Puuska, H.-M. (2010). Effects of scholar's gender and professional position on publishing productivity in different publication types: Analysis of a Finnish university. *Scientometrics, 82*(2), 419–437.
Ruiz-Castillo, J., & Costas, R. (2014). The skewness of scientific productivity. *Journal of Informetrics, 8*(4), 917–934.
Salganik, M. J. (2018). *Bit by bit. Social research in a digital age*. Princeton University Press.
Selwyn, N. (2019). *What is digital sociology?* Polity Press.
Stephan, P. (2012). *How economics shapes science*. Harvard University Press.
Stephan, P. E., & Levin, S. G. (1992). *Striking the mother lode in science: The importance of age, place, and time*. Oxford University Press.
Sugimoto, C., & Larivière, V. (2018). *Measuring research: What everyone needs to know*. Oxford University Press.
Taylor, B. J., & Cantwell, B. (2019). *Unequal higher education: Wealth, status, and student opportunity*. Rutgers University Press.
Weinberger, M., & Zhitomirsky-Geffet, M. (2021). Diversity of success: Measuring the scholarly performance diversity of tenured professors in Israeli academia. *Scientometrics, 126*, 2931–2970.
Xie, Y. (2014). 'Undemocracy': Inequalities in science. *Science, 344*(6186), 809–810.
Yair, G., Gueta, N., & Davidovitch, N. (2017). The law of limited excellence: Publication productivity of Israel Prize laureates in the life and exact sciences. *Scientometrics, 113*(1), 299–311.
Yin, Z., & Zhi, Q. (2017). Dancing with the academic elite: A promotion or hindrance of research production? *Scientometrics, 110*(1), 17–41.
Yuret, T. (2018). Path to success: An analysis of US educated elite academics in the United States. *Scientometrics, 117*, 105–121.
Zuckerman, H. (1988). The sociology of science. In N. J. Smelser (Ed.), *Handbook of sociology* (pp. 511–574). Sage.

**Publisher's note**  Springer Nature remains neutral with regard to jurisdictional claims in published maps and institutional affiliations.

## Authors and Affiliations

**Marek Kwiek**[1,2,3] · **Wojciech Roszka**[2,4]

Wojciech Roszka
wojciech.roszka@ue.poznan.pl

1. Institute for Advanced Studies in Social Sciences and Humanities, Adam Mickiewicz University of Poznan, Poznan, Poland
2. Center for Public Policy Studies, Adam Mickiewicz University, Poznan, Poland
3. German Centre for Higher Education Research and Science Studies (DZHW), Berlin, Germany
4. Poznan University of Economics and Business, Poznan, Poland